\documentclass[showpacs,prd,preprintnumbers,nofootinbib]{revtex4}%
\usepackage{amssymb}
\usepackage[T1]{fontenc}
\usepackage{amsmath}
\usepackage{graphicx}
\usepackage{hyperref}
\usepackage{color}
\usepackage{amsfonts}%
\providecommand{\U}[1]{\protect\rule{.1in}{.1in}}
\begin{document}
\title{Nuclei as near BPS-Skyrmions}
\author{Eric Bonenfant and Luc Marleau}
\affiliation{D\'{e}partement de Physique, de G\'{e}nie Physique et d'Optique,
Universit\'{e} Laval, Qu\'{e}bec, Qu\'{e}bec, Canada G1K 7P4}
\date{\today}

\begin{abstract}
We study a generalization of the Skyrme model with the inclusion of a
sixth-order term and a generalized mass term. We first analyze the model in a
regime where the nonlinear $\sigma$ and Skyrme terms are switched to zero
which leads to well-behaved analytical BPS-type solutions. Adding
contributions from the rotational energy, we reproduce the mass of the most
abundant isotopes to rather good accuracy. These BPS-type solutions are then
used to compute the contributions from the nonlinear sigma and Skyrme terms
when these are switched on. We then adjust the four parameters of the model
using two different procedures and find that the additional terms only
represent small perturbations to the system. We finally calculate the binding
energy per nucleon and compare our results with the experimental values.

\end{abstract}

\pacs{12.39.Dc, 11.10.Lm}
\maketitle

\section{\label{sec:Intro}Introduction}

The Skyrme model \cite{Skyrme} is nowadays one of the strongest candidates for
a description of the low-energy regime of QCD. Developed in the beginning of
the 60's by T.H.R. Skyrme, it consists of a nonlinear theory of mesons where
its main feature is the presence of topological solitons as solutions. Each of
these solutions is associated with a conserved topological charge, the winding
number, which Skyrme interpreted as the baryon number, thus leading him to
state that the solitons are baryons emerging from a meson field. The $\frac
{1}{N_{C}}$ expansion of QCD introduced by t'Hooft \cite{thooft} in the
mid-70's and the later connection from Witten \cite{Witten} with the model
developed by Skyrme brought some support to this interpretation.

Since its original formulation, the Skyrme model has been able to predict the
properties of the nucleon within a precision of 30\%. Several modifications to
the model have been considered to improve these predictions, from the
generalization of the mass term \cite{Marleau1,KPZ,Marleau2} to the explicit
addition of vector mesons \cite{Sutcliffe, Adkins2}, aside form higher order
terms in derivatives of the pion field \cite{Marleau1}. Unfortunately, the
analysis of these models has been hampered by their nonlinear nature and the
absence of analytical solutions. Indeed, all the solutions rely on numerical
computation at some point whether one uses the rational map ansatz
\cite{Houghton}, which turns out to be a rather good approximation of the
angular dependance, or a full fledge numerical algorithm like simulated
annealing \cite{Marleau3,Houghton2} to find an exact solution of the energy
functional. Clearly, even a prototype model with analytical solutions would
allow to go deeper in the investigation of the properties and perhaps identify
novel features of the Skyrmions.

In a recent study, Adam, Sanchez-Guillen and Wereszczynski (ASW) \cite{Adam}
obtained an analytical solution by considering a model consisting only of a
term of order six in derivatives and a potential which correspond to the
customary mass term for pions in the Skyrme model \cite{Adkins}. Their
calculations lead to a compacton-type solution with size growing as
$n^{\frac{1}{3}}$, with $n$ is the winding number, a result in general
agreement with experimental observations. Another important remark on their
study is that their solutions are of BPS-type, i.e. they saturate a
Bogomolny's bound. Even though physical nuclei do not saturate such a bound,
the small value of the binding energy may be one of the motivation for
solution of this type. Let us also mention that recently Sutcliffe
\cite{Sutcliffe2} found that BPS-type Skyrmions also emerge from models when a
large number of vector mesons are added to the Skyrme model. However, the
analysis of ASW neglects rotational or isorotational energies of nuclei and
perhaps the oddest feature of the model is that it does not contain any of the
terms that Skyrme originally introduced in his model, the nonlinear sigma and
so-called Skyrme terms which are of order 2 and 4 in derivatives respectively.
Being a effective theory of QCD, there is nevertheless no reason to omit such
contributions. In their work ASW further suggest that their analytical
solutions found could be used to compute the contributions from the terms of
the original Skyrme lagrangian assuming they are small and do not affect
significantly the overall solutions. Unfortunately the nature of the solution
leads to singularities in the computation of the energies related to the
nonlinear sigma and Skyrme terms.

In this work, we find analytical BPS-type solutions for a Lagrangian similar
to the one in \cite{Adam} which allows to consider contributions from the
original Skyrme Lagrangian as small perturbations. The analysis also includes
contributions for (iso)rotational energies providing a more realistic
description of nuclei. The paper is divided as follows: In section
\ref{sec:Skyrme}, we introduce the general form of this generalized Skyrme
model and find expressions for the static energies. Next, we quantify
semiclassically the zero modes of the Skyrmions, which will allow to compute
rotational contributions to the total energy coming from the spin as well as
the isospin of the nuclei. In section \ref{sec:Model}, we choose an adequate
potential (or mass term) and switch off the nonlinear sigma and Skyrme terms.
We then find a simple analytical form of the BPS-type solutions for the
remaining Lagrangian. It turns out that all the properties of the nuclei can
be calculated analytically. In section \ref{sec:Approximation}, we use the
solution to compute the properties of the full Lagrangian. Fitting the
different parameters of the model with nuclear mass data \cite{Nucltable} we
verify that the contributions from the nonlinear sigma and Skyrme terms remain
small and that the analytical solution is a good approximation.

\section{\label{sec:Skyrme}Lagrangian of the Skyrme model}

The model proposed by ASW is based on the Lagrangian density%
\begin{equation}
\mathcal{L}=\mathcal{L}_{6}-\mu^{2}V=-\frac{3}{2}\frac{\lambda^{2}}{16^{2}%
}\text{Tr}\left(  \left[  L_{\mu},L_{\nu}\right]  \left[  L^{\nu},L^{\lambda
}\right]  \left[  L_{\lambda},L_{\mu}\right]  \right)  -\frac{1}{2}\mu
^{2}\text{Tr}\left[  1-U\right]  . \label{L_adam}%
\end{equation}
where $U=\phi_{0}+i\tau_{i}\phi_{i}$ is the $SU(2)$ matrix representing the
meson fields and $L_{\mu}=U^{\dagger}\partial_{\mu}U$ is the left-handed
current. The model leads to BPS-type solitons. The constants $\lambda$ and
$\mu$ are the only free parameters of the model with units MeV$^{-1}$ and
MeV$^{2}$ respectively. Using scaling arguments, one can show that the term of
order 6 in field derivatives, $\mathcal{L}_{6},$ prevents the soliton from
shrinking to zero size while the second term, often called the mass term,
stabilize the solution against arbitrary expansion.

On the other hand, the original Skyrme model consists of the two completely
different terms
\begin{equation}
\label{termes}\mathcal{L}=\mathcal{L}_{nl\sigma}+\mathcal{L}_{Sk}%
\end{equation}
with%
\begin{equation}
\mathcal{L}_{nl\sigma}=\alpha\text{Tr}\left[  L_{\mu}L^{\mu}\right]
,\qquad\mathcal{L}_{Sk}=\beta\text{Tr}\left(  \left[  L_{\mu},L_{\nu}\right]
^{2}\right)
\end{equation}
the nonlinear sigma and so-called Skyrme terms which are of order 2 and 4 in
derivatives respectively. Here $\left[  \alpha\right]  =\text{MeV}^{2}$ and
$\beta$ is a dimensionless constant.

We shall consider here a model containing the four terms i.e. an extension of
the Skyrme model with a sixth order term in derivatives and generalized mass
term. The Lagrangian density reads
\begin{equation}
\mathcal{L}=-\mu^{2}V(U)-\alpha\text{Tr}\left[  L_{\mu}L^{\mu}\right]
+\beta\text{Tr}\left(  \left[  L_{\mu},L_{\nu}\right]  ^{2}\right)  -\frac
{3}{2}\frac{\lambda^{2}}{16^{2}}\text{Tr}\left(  \left[  L_{\mu},L_{\nu
}\right]  \left[  L^{\nu},L^{\lambda}\right]  \left[  L_{\lambda},L_{\mu
}\right]  \right)  \label{L_S}%
\end{equation}
We are interested in the regime where $\alpha$ and $\beta$ are small so that
$\mathcal{L}_{nl\sigma}$ and $\mathcal{L}_{Sk}$ can be considered as small
perturbations to (\ref{L_adam}). Usually, the potential $V(U)$ is chosen such
that it reproduces the mass term for pions when small fluctuations of the
fields are considered
\begin{equation}
U=e^{2i\tau_{i}\frac{\pi_{i}}{F_{\pi}}}\sim1+2i\tau_{i}\frac{\pi_{i}}{F_{\pi}%
}.
\end{equation}
where $F_{\pi}=4\sqrt{\alpha}$ is the pion decay constant. Since $U$ is an
$SU\left(  2\right)  $ matrix, the meson fields obey the condition
\begin{equation}
\phi_{0}^{2}+\phi_{i}^{2}=1
\end{equation}
to limit the degrees of freedom to three. The boundary condition
\begin{equation}
U\left(  r\rightarrow\infty\right)  =\mathcal{I}_{2\times2}%
\end{equation}
with $\mathcal{I}_{2\times2}$ the two-dimensional unit matrix, ensures that
each solution for the Skyrme field falls into a topological sector
characterized by a conserved topological charge
\begin{equation}
B=-\frac{\epsilon^{ijk}}{48\pi^{2}}\int d^{3}x\text{Tr}\left(  L_{i}\left[
L_{j},L_{k}\right]  \right)  .
\end{equation}
The static energy can then be calculated using
\begin{equation}
E_{stat}=-\int d^{3}x\mathcal{L}_{stat}.
\end{equation}
We may conveniently write a general solution as%
\begin{equation}
U=e^{i\mathbf{n}\cdot\mathbf{\tau}F}=\cos F+i\mathbf{n}\cdot\mathbf{\tau}\sin
F \label{Hedgehog}%
\end{equation}
where $\mathbf{\hat{n}}$ is the unit vector
\[
\mathbf{\hat{n}}=\left(  \sin\Theta\cos\Phi,\sin\Theta\sin\Phi,\cos
\Theta\right)
\]
or
\[
\phi_{a}=\left(  \cos F,\sin F\sin\Theta\cos\Phi,\sin F\sin\Theta\sin\Phi,\sin
F\cos\Theta\right)  .
\]
Following ASW \cite{Adam}, we consider solutions of the form that saturates
the Bogomolny's bound for (\ref{L_adam})%
\begin{equation}
F=F(r),\qquad\Theta=\theta,\qquad\Phi=n\phi
\end{equation}
where $n$ is an integer. The static energy (\ref{L_S}) becomes
\begin{align}
E_{stat}  &  =-\int d^{3}x\mathcal{L}_{stat}=4\pi\int r^{2}dr\left(  \mu
^{2}V+\frac{9\lambda^{2}}{16}n^{2}F^{\prime}{}^{2}\frac{\sin^{4}F}{r^{4}%
}\right. \label{Estat}\\
&  +\left.  2\alpha\left[  F^{\prime2}+\left(  n^{2}+1\right)  \frac{\sin
^{2}F}{r^{2}}\right]  +16\beta\frac{\sin^{2}F}{r^{2}}\left[  \left(
n^{2}+1\right)  F^{\prime2}+n^{2}\frac{\sin^{2}F}{r^{2}}\right]  \right)
\nonumber
\end{align}
where $F^{\prime}=\partial F/\partial r$ and the topological charge is simply
$B=n$.

In order to represent physical nuclei, we have to quantize the solitons using
a semiclassical method described in the next section. Using the appropriate
spin and isospin numbers, we will then be able to calculate the total energy
for each nuclei.

\section{\label{sec:Quantization}Quantization}

Because the topological solitons occupy a spatial volume that is nonzero,
usual quantization procedures are no longer available. We therefore have to
use a semiclassical quantization method by adding an explicit time dependence
to the zero modes of the Skyrmion. Performing time-dependent (iso)rotations on
the Skyrme field by $SU(2)$ matrix $A(t)$ and $B(t)$ yield
\begin{equation}
\tilde{U}\left(  \mathbf{r},t\right)  =A(t)U\left(  R(B(t))\mathbf{r}\right)
A(t)
\end{equation}
where $R_{ij}(B(t))=\frac{1}{2}\text{Tr}\left[  \tau_{i}B\tau_{j}B^{\dagger
}\right]  $ is the associated $SO(3)$ rotation matrix. Upon insertion of this
ansatz in the time-dependent part of (\ref{L_S}), we write the rotational
lagrangian as
\begin{equation}
\mathcal{L}_{rot}=\frac{1}{2}a_{i}U_{ij}a_{j}-a_{i}W_{ij}b_{j}+\frac{1}%
{2}b_{i}V_{ij}b_{j}%
\end{equation}
with $U_{ij}$, $V_{ij}$ and $W_{ij}$ the inertia tensors
\begin{align}
U_{ij}  &  =-\int\text{d}^{3}x\biggl\{2\alpha\text{Tr}\left(  T_{i}%
T_{j}\right)  +4\beta\text{Tr}\left(  \left[  L_{k},T_{i}\right]  \left[
L_{k},T_{j}\right]  \right)  \biggr.\nonumber\\
&  \biggl.+\frac{9\lambda^{2}}{16^{2}}\text{Tr}\left(  \left[  T_{i}%
,L_{k}\right]  \left[  L_{k},L_{n}\right]  \left[  L_{n},T_{j}\right]
\right)  \biggr\}, \label{eq:Uij}%
\end{align}%
\begin{align}
V_{ij}  &  =-\epsilon_{ikl}\epsilon_{jmn}\int\text{d}^{3}xx_{k}x_{m}%
\biggl\{2\alpha\text{Tr}\left(  L_{l}L_{n}\right)  +4\beta\text{Tr}\left(
\left[  L_{p},L_{l}\right]  \left[  L_{p},L_{n}\right]  \right)
\biggr.\nonumber\\
&  \biggl.+\frac{9\lambda^{2}}{16^{2}}\text{Tr}\left(  \left[  L_{l}%
,L_{p}\right]  \left[  L_{p},L_{q}\right]  \left[  L_{q},L_{n}\right]
\right)  \biggr\}, \label{eq:Vij}%
\end{align}%
\begin{align}
W_{ij}  &  =\epsilon_{jkl}\int\text{d}^{3}xx_{k}\biggl\{2\alpha\text{Tr}%
\left(  T_{i}L_{l}\right)  +4\beta\text{Tr}\left(  \left[  L_{p},T_{j}\right]
\left[  L_{p},L_{n}\right]  \right)  \biggr.\nonumber\\
&  \biggr.+\frac{9\lambda^{2}}{16^{2}}\text{Tr}\left(  \left[  T_{i}%
,L_{m}\right]  \left[  L_{m},L_{n}\right]  \left[  L_{n},L_{l}\right]
\right)  \biggr\} \label{eq:Wij}%
\end{align}
and $T_{i}=iU^{\dagger}\left[  \frac{\tau_{i}}{2},U\right]  $. Assuming a
solution of the form (\ref{Hedgehog}), the inertia tensors becomes all
diagonal and furthermore, one can show that $U_{11}=U_{22}\neq U_{33}$ with
similar identities for the $V_{ij}$ and $W_{ij}$ tensors. Finally the general
expressions for the moments of inertia coming from each pieces of the
Lagrangian read%

\begin{align}
U_{11}  &  =\frac{4\pi}{3}\int r^{2}dr\sin^{2}F\left(  8\alpha+16\beta\left[
4F^{\prime2}+\left(  3n^{2}+1\right)  \frac{\sin^{2}F}{r^{2}}\right]
+\frac{9\lambda^{2}}{4}\frac{\left(  3n^{2}+1\right)  }{4}F^{\prime2}%
\frac{\sin^{2}F}{r^{2}}\right)  ,\label{U11}\\
V_{11}  &  =\frac{4\pi}{3}\int r^{2}dr\sin^{2}F\left(  2\left(  n^{2}%
+3\right)  \alpha+16\beta\left[  \left(  n^{2}+3\right)  F^{\prime2}%
+4n^{2}\frac{\sin^{2}F}{r^{2}}\right]  +\frac{9\lambda^{2}}{4}n^{2}F^{\prime
2}\frac{\sin^{2}F}{r^{2}}\right)  , \label{V11}%
\end{align}
and the general expression for $U_{33}$ can be obtained by setting $n=1$ in
the integrand of (\ref{U11}). It turns out that expressions (\ref{eq:Uij}%
)-(\ref{eq:Wij}) leads to $W_{11}=W_{22}=0$ for $\left\vert n\right\vert
\geq2$ and $n^{2}U_{33}=nW_{33}=V_{33}$. Otherwise, for $\left\vert
n\right\vert =1$ we have spherical symmetry and,
\begin{equation}
W_{11}=\frac{4\pi}{3}\int r^{2}dr\sin^{2}F\left(  8\alpha+16\beta\left(
4\frac{\sin^{2}F}{r^{2}}+4F^{\prime}\right)  +\frac{9\lambda^{2}}{4}%
F^{\prime2}\frac{\sin^{2}F}{r^{2}}\right)  .
\end{equation}

Following Houghton and Magee \cite{Houghton2}, we now write the rotational
hamiltonian as
\begin{equation}
H_{rot}=\frac{1}{2}\left[  \frac{\left(  L_{1}+W_{11}\frac{K_{1}}{U_{11}%
}\right)  ^{2}}{V_{11}-\frac{W_{11}^{2}}{U_{11}}}+\frac{\left(  L_{2}%
+W_{22}\frac{K_{2}}{U_{22}}\right)  ^{2}}{V_{22}-\frac{W_{22}^{2}}{U_{22}}%
}+\frac{\left(  L_{3}+W_{33}\frac{K_{3}}{U_{33}}\right)  ^{2}}{V_{33}%
-\frac{W_{33}^{2}}{U_{33}}}+\frac{K_{1}^{2}}{U_{11}}+\frac{K_{2}^{2}}{U_{22}%
}+\frac{K_{3}^{2}}{U_{33}}\right]  \label{Hrot}%
\end{equation}
with ($K_{i}$) $L_{i}$ the body-fixed (iso)rotation momentum canonically
conjugate to $a_{i}$ and $b_{i}$ respectively. The expression for the
rotational energy of the nucleon has been obtained in \cite{Houghton2} and
reads, for a spherical symmetry
\begin{equation}
E_{rot}^{N}=\frac{3}{8U_{11}}.
\end{equation}
For the deuteron, the rotational energy has been calculated assuming an axial
symmetric solution \cite{Marleau4}
\begin{equation}
E_{rot}^{D}=\frac{1}{2V_{11}}+\frac{1}{2V_{22}}%
\end{equation}
which reduces to
\begin{equation}
E_{rot}^{D}=\frac{1}{V_{11}}%
\end{equation}
for the axial ansatz (\ref{Hedgehog}). It is easy to calculate the rotational
energies for nuclei with winding number $\left\vert n\right\vert \geq3$. The
axial symmetry of the solution imposes the constraint $L_{3}+nK_{3}=0$ which
is simply the statement that a spatial rotation by an angle $\theta$ about the
axis of symmetry can compensated by an isorotation of $-n\theta$ about the
$\tau_{3}$ axis. It also implies that$\ n^{2}U_{33}=nW_{33}=V_{33}$. Recalling
that $W_{11}=W_{22}=0$ for these values of $n$, the rotational hamiltonian
reduces to%
\begin{equation}
H_{rot}=\frac{1}{2}\left[  \frac{\mathbf{L}^{2}}{V_{11}}+\frac{\mathbf{K}^{2}%
}{U_{11}}+\left(  \frac{1}{U_{33}}-\frac{1}{U_{11}}-\frac{n^{2}}{V_{11}%
}\right)  K_{3}^{2}\right]
\end{equation}
These momenta are related to the usual space-fixed isospin ($\mathbf{I}$) and
spin ($\mathbf{J}$) by the orthogonal transformations
\begin{equation}
I_{i}=-R(A_{1})_{ij}K_{j}, \label{eq:I}%
\end{equation}%
\begin{equation}
J_{i}=-R(A_{2})_{ij}^{\text{T}}L_{j}. \label{eq:J}%
\end{equation}
According to (\ref{eq:I}) and (\ref{eq:J}), we see that the Casimir invariants
satisfy $\mathbf{K}^{2}=\mathbf{I}^{2}=I(I+1)$ and $\mathbf{L}^{2}%
=\mathbf{J}^{2}=J(J+1)$ so the rotational hamiltonian is given by
\begin{equation}
H_{rot}=\frac{1}{2}\left[  \frac{J(J+1)}{V_{11}}+\frac{I(I+1)}{U_{11}}+\left(
\frac{1}{U_{33}}-\frac{1}{U_{11}}-\frac{n^{2}}{V_{11}}\right)  K_{3}%
^{2}\right]  . \label{Erot}%
\end{equation}

\section{\label{sec:Model}BPS-type solutions}

Let's consider a model similar to \cite{Adam} composed of the term of order
six in derivatives plus a potential by setting $\alpha$, $\beta=0$
\begin{equation}
\mathcal{L}=-\frac{3}{2}\frac{\lambda^{2}}{16^{2}}\text{Tr}\left(  \left[
L_{\mu},L_{\nu}\right]  \left[  L^{\nu},L^{\lambda}\right]  \left[
L_{\lambda},L_{\mu}\right]  \right)  -\mu^{2}V(U).
\end{equation}
Using the results of section \ref{sec:Skyrme}, the static energy is
\begin{equation}
E_{stat}=4\pi\int dr\left(  \frac{9\lambda^{2}n^{2}}{4}\frac{\sin^{4}F}%
{4r^{2}}F^{\prime2}+\mu^{2}V(U)\right)  .
\end{equation}
The minimization of the static energy of the soliton, leads to the
differential equation for $F$ {}
\begin{equation}
\frac{9\lambda^{2}n^{2}}{4}\frac{\sin^{2}F}{2r^{2}}\partial_{r}\left(
\frac{\sin^{2}F}{r^{2}}F^{\prime}\right)  -\mu^{2}V_{F}=0.
\label{minimisation}%
\end{equation}
A change of variable $z=\frac{2\sqrt{2}\mu r^{3}}{9n\lambda}$ allows to
rewrite (\ref{minimisation}) in a simple form
\begin{equation}
\sin^{2}F\partial_{z}\left[  \sin^{2}F\left(  \partial_{z}F\right)  \right]
-\mu^{2}\frac{\partial V}{\partial F}=0.
\end{equation}
This last equation can be integrated
\begin{equation}
\frac{1}{2}\sin^{4}F\left(  F_{z}\right)  ^{2}=V \label{equipartition}%
\end{equation}
and inserting the expression for $z$, provide an expression which amounts to a
statement of equipartition of the energy, i.e. the term of order 6 in
derivatives and the potential contribute equally to the total energy. ASW has
shown that a solution of (\ref{equipartition}) saturates the Bogomolny's bound
\cite{Adam}. From (\ref{equipartition}) we obtain the following useful
relation between the function $F$ and the potential
\begin{equation}
\int dF\frac{\sin^{2}F}{\sqrt{2V}}=\pm\left(  z-z_{0}\right)  \label{useful}%
\end{equation}
with $z_{0}$ an integration constant.

Now comes the time to choose a specific potential. The\ choice for mass term
of the Skyrme is not unique and indeed has been the object of several
discussions \cite{Adkins,Marleau1,KPZ}. The usual mass term $V=1-\cos F$ was
considered in \cite{Adam}. Solving (\ref{useful}) for $F$
\begin{equation}
F(r)=\left\{
\begin{tabular}
[c]{lll}%
$2\arccos\left(  \nu r^{3}\right)  $ & $\qquad\text{for}$ & $r\in\left[
0,\nu^{-\frac{1}{3}}\right]  $\\
$0$ & $\qquad\text{for}$ & $r\geq\nu^{-\frac{1}{3}}$%
\end{tabular}
\text{ }\right.  \label{adamsoln}%
\end{equation}
where $\nu=\frac{\mu}{18n\lambda}$ is a constant depending on the parameters
$\lambda$, $\mu$ and $n$. Note that $F^{\prime}$ diverges as $\ r\rightarrow
\nu^{-\frac{1}{3}}.$ Since this solution saturates the Bogomolny's bound, the
static energy is proportional to the baryon number $B=n$ .

A question arises as how would the nonlinear $\sigma$ and Skyrme term affect
the energy of such Skyrmions. Switching them on slowly by moving $\alpha$ and
$\beta$ away from zero could give an estimate of their contributions.
Unfortunately, it turns out that simply substituting the solution
(\ref{adamsoln}) in the expression for energy associated with the full
Lagrangian leads to divergences. So, however small the parameters $\alpha$ and
$\beta$ are, these BPS solutions cannot be considered as appropriate
approximations of the solutions for (\ref{L_S}).

Yet it could be interesting to analyze the full Lagrangian (\ref{L_S}) in a
regime close to a BPS Skyrmion. For this purpose, we propose to write\ the
potential in the form of the generalized mass term introduced by
\cite{Marleau1}
\begin{align}
V  &  =-\sum_{k=1}^{\infty}C_{k}\text{Tr}\left[  U^{k}+U^{\dagger k}-2\right]
\nonumber\\
&  =-4\pi\sum_{k=1}^{\infty}\int r^{2}dr8C_{k}\sin^{2}\left(  \frac{k\xi}%
{2}\right)  .
\end{align}
The main motivation for this choice is that the potential can be written in a
simple form in terms of pion fields. Furthermore this particular framework
insures that one recovers the chiral symmetry breaking pion mass term
$-\frac{1}{2}m_{\pi}^{2}\mathbf{\pi}\cdot\mathbf{\pi}$ \ in the limit of small
pion field fluctuations provided
\begin{equation}
\sum_{k=1}^{\infty}k^{2}C_{k}=-\frac{m_{\pi}^{2}F_{\pi}^{2}}{16}.
\end{equation}

For practical purposes, one requires (i) an expression of the potential that
is simple enough to allow the analytical integration of the left-hand side of
equation (\ref{useful}), (ii) that the results leads to an invertible function
to be able to write the chiral profile $F$ as a function of $r$ and finally
(iii) that $F\left(  r\right)  $ is well behaved$.$ A most convenient choice
is
\begin{equation}
V=\sin^{2}\left(  \frac{F}{2}\right)  \cos^{6}\left(  \frac{F}{2}\right)  .
\label{potential}%
\end{equation}
Expanding the expression (\ref{potential}), the coefficients $C_{k}$ are
\begin{equation}
C_{1}=-\frac{\mu^{2}}{128},\qquad C_{2}=\frac{\mu^{2}}{128},\qquad C_{3}%
=\frac{\mu^{2}}{128},\qquad C_{4}=\frac{\mu^{2}}{512},\qquad C_{k>4}=0.
\end{equation}
Integrating (\ref{useful}) we get the general solution
\begin{equation}
F(r)=2\arccos\left(  e^{\pm\nu\left(  r^{3}-r_{0}^{3}\right)  }\right)
\label{Fsoln}%
\end{equation}
with $\nu=\frac{\mu}{18n\lambda}$. In order that the baryon number corresponds
to $\left\vert B\right\vert =n,$ one must require that
\[
F(\infty)-F(0)=\mp\pi
\]
for $B$ positive or negative respectively. Accordingly we choose the boundary
conditions $F(0)=0$ and $F(\infty)=\mp\pi$ which sets the integration constant
$r_{0}$ to zero and allows to write%
\begin{equation}
F(r)=\mp2\left\vert \arccos\left(  e^{-\nu r^{3}}\right)  \right\vert
\end{equation}
where we use the absolute value to dispose of the sign ambiguity of the arccos
function. Note that here contrary to \cite{Adam} we do not get a compacton
type solution but a well behaved function with a continuous first derivative.
All calculations regarding energy can be performed analytically, i.e. static
energy and the moments. For example, the baryon density is given by the radial
function%
\[
B(r)=\frac{2\mu}{3\pi^{2}\lambda}e^{-\frac{\mu r^{3}}{6n\lambda}}\left(
1-e^{-\frac{\mu r^{3}}{9n\lambda}}\right)
\]
which upon integration leads to baryon number $B=n.$ Experimentally the size
of the nucleus is known to behave as
\[
R=R_{0}B^{\frac{1}{3}}=\left(  1.25\text{ fm}\right)  B^{\frac{1}{3}}%
\]
where $R_{0}=1.25$ fm$.$ It is interesting to note that the baryon number
distribution is zero at $r=0$ but has maximum value $\frac{\sqrt{3}\mu}%
{8\pi^{2}\lambda}$ independent of $n$ which is positioned at
\begin{equation}
r_{\max}=\left(  \frac{9\lambda}{\mu}\ln\left(  \frac{4}{3}\right)  \right)
^{\frac{1}{3}}B^{\frac{1}{3}} \label{rmax}%
\end{equation}
where $r_{\max}$ here is in units of MeV$^{-1}$. Accordingly the size of the
nucleus $r_{\max}$ \ is proportional to $B^{\frac{1}{3}}$ with $R_{0}$
depending only on the ratio $\lambda/\mu$. Similarly expressions can be
obtained for energy and moment of inertia density. Using (\ref{Fsoln}), they
yield
\begin{align}
E  &  =2n\pi\mu\lambda\nonumber\\
V_{11}  &  =n^{2}U_{33}=\frac{4n^{2}}{\left(  3n^{2}+1\right)  }U_{11}%
=2\pi\left(  \frac{\lambda n}{3\mu}\right)  ^{\frac{5}{3}}\mu^{2}\Gamma\left(
\frac{2}{3}\right)  \left(  16\cdot3^{\frac{1}{3}}-9\cdot2^{\frac{2}{3}%
}\right)
\end{align}
where $\Gamma\left(  x\right)  $ is the gamma function. Combining these
results in (\ref{Erot})%

\begin{equation}
H_{rot}=\frac{1}{2U_{11}}\left[  J(J+1)\frac{\left(  3n^{2}+1\right)  }%
{4n^{2}}+I(I+1)-K_{3}^{2}\right]  .
\end{equation}
Note that this last result only hold for $\alpha=\beta=0$ and the solution
(\ref{Fsoln}). The last term is either zero or negative. Depending on the
dimension of the spin and isospin representation, the diagonalization of this
hamiltonian will lead to a number of eigenstates. We are interested in the
lowest eigenvalue of $H_{rot}$ which points towards the eigenstate $\left\vert
i,i_{3},k_{3}\right\rangle \left\vert j,j_{3},l_{3}\right\rangle $ with the
largest possible eigenvalue $k_{3}.$ Since $\mathbf{K}^{2}=\mathbf{I}^{2}$ and
$\mathbf{L}^{2}=\mathbf{J}^{2},$ the state with highest weight is
characterized by $k_{3}=I$ and $l_{3}=j$ and since since nuclei are build of
$B$ fermions $j\leq B/2.$ On the other hand the axial symmetry of the
solutions implies that $k_{3}=-l_{3}/n.$ We recall that these solutions are
approximations. Then for even $B$ nuclei, the integer part of $\left\vert
l_{3}/n\right\vert $ is
\[
0\leq\left\vert k_{3}\right\vert =\left[  \left\vert \frac{l_{3}}%
{n}\right\vert \right]  \leq\left[  \left\vert \frac{j}{n}\right\vert \right]
\leq\left[  \left\vert \frac{B}{2n}\right\vert \right]  =0
\]
so it leads to $\left\vert k_{3}\right\vert =0$. Similarly for half-integer
spin nuclei,
\[
\frac{1}{2}\leq\left\vert k_{3}\right\vert \leq\left\vert \frac{j}%
{n}\right\vert \leq\left\vert \frac{B}{2n}\right\vert =\frac{1}{2}%
\]
So we shall assume for simplicity%
\[
\kappa=\max(\left\vert k_{3}\right\vert )=\left\{
\begin{tabular}
[c]{l}%
$0\qquad$for $B=$ even\\
$\frac{1}{2}\qquad$for $B=$ odd
\end{tabular}
\ \right.
\]
The rotational energy is given by%
\begin{equation}
E_{rot}=\frac{1}{2U_{11}}\left[  j(j+1)\frac{\left(  3n^{2}+1\right)  }%
{4n^{2}}+i(i+1)-\kappa^{2}\right]  . \label{Ekappa}%
\end{equation}

It remains to fix the values of the parameters $\lambda$ and $\mu$. In order
to do so, we choose as input parameters the experimental mass of the nucleon
and for simplicity, a nucleus $\ X$ with zero (iso)rotational energy (i.e. a
nucleus with zero spin and isospin). The total energy of these two states are
\begin{align}
E_{N}  &  =\left.  \left(  E+\frac{3}{8U_{11}}\right)  \right\vert
_{n=1}\nonumber\\
&  =2\pi\mu\lambda+\frac{1}{\mu^{2}}\left(  \frac{3\mu}{\lambda}\right)
^{\frac{5}{3}}\frac{3}{16\pi\Gamma\left(  \frac{2}{3}\right)  \left(
16\cdot3^{\frac{1}{3}}-9\cdot2^{\frac{2}{3}}\right)  }\\
E_{X}  &  =\left.  E\right\vert _{n=B}=2B\pi\mu\lambda
\end{align}
Solving for $\lambda$ and $\mu$ we get%
\begin{align}
\lambda &  =\frac{3\cdot3^{\frac{1}{4}}}{\left(  E_{X}\right)  ^{\frac{1}{4}%
}\sqrt{\pi}\left(  \left(  E_{X}-nE_{N}\right)  \left(  9\cdot2^{\frac{2}{3}%
}-16\cdot3^{\frac{1}{3}}\right)  \Gamma\left(  \frac{5}{3}\right)  \right)
^{\frac{3}{4}}}\nonumber\\
\mu &  =\frac{\left(  E_{X}\right)  ^{\frac{5}{4}}\left(  \left(  E_{X}%
-nE_{N}\right)  \left(  9\cdot2^{\frac{2}{3}}-16\cdot3^{\frac{1}{3}}\right)
\Gamma\left(  \frac{5}{3}\right)  \right)  ^{\frac{3}{4}}}{24\cdot3^{\frac
{1}{4}}\sqrt{\pi}} \label{sollambdamu}%
\end{align}
As an example, we choose the nucleus $X$ to be the helium-4, the first doubly
magic number nucleus. The mass of the helium-4 nucleus has no (iso)rotational
parts since it has zero spin and isospin. Setting the mass of the nucleon as
the average mass of the proton and neutron i.e. $E_{N}=938.919$ MeV and the
mass of the helium nucleus to $E_{He}=3727.38$ MeV, we obtain the numerical
value $\lambda=0.00491505$ MeV$^{-1}$ and $\mu=30174.2$ MeV$^{2}$. We shall
refer to this set of parameters as Set Ia.

Experimentally the size of the nucleus is known to behave as
\[
R=R_{0}B^{\frac{1}{3}}%
\]
with $R_{0}=1.25$ fm We get a similar behaviour for $r_{\max}$ in
(\ref{rmax}).%
\begin{equation}
r_{\max}=\left(  1.4798\text{ fm}\right)  B^{\frac{1}{3}}%
\end{equation}

Combining (\ref{sollambdamu}) with (\ref{Estat}) and (\ref{Ekappa}), the mass
of any nucleus can be expressed as a analytical function of the input
parameters $E_{N}$ and $E_{He}$. In general it depends on the baryon number as
well as the spin and the isospin of the isotope. The spin of the most abondant
isotopes are known. The isospins are not so well known so we resort to the
usual assumption that the most abundant isotopes correspond to states with
lowest isorotational energy, i.e. states where the isospin $I$ has the lowest
value that the conservation of the third component of isospin $I_{3}$ allows.
Accordingly,
\begin{align}
I  &  =\left\vert I_{3}\right\vert =\frac{1}{2}\left\vert \#\text{ of proton
}-\#\text{ of neutron}\right\vert \nonumber\\
&  =\left\vert \frac{A}{2}-Z\right\vert \label{Isospin}%
\end{align}
Table I shows the results for the a few isotopes. The resulting predictions
are accurate to $0.3\%$ or better even for heavier nuclei which is rather
surprising since the model involves only to two free parameters $\lambda$ and
$\mu.$

The computation were repeated using as input parameter $X = $ $^{16}$O and
$^{40}$Ca, two other doubly magic nuclei (also shown in Table I, Set Ib and
Set Ic respectively). These set the parameters to $\lambda=0.00449295$
MeV$^{-1}$ and $\mu=32977.0$ MeV$^{2}$ and \ to $\lambda=0.00426504$
MeV$^{-1}$ and $\mu=34717.8$ MeV$^{2}$ respectively. Using these heavier
elements as input parameters changes slightly the overall predicting accuracy.
Whereas the best overall accuracy is achieved using $^{16}$O parametrization
in Set Ib, the lightest isotopes are best described by choosing $^{4}$He as
input (Set Ia). Note that the lightest nuclei have lower moments of inertia
and get relatively large rotational contribution to their mass. Consequently
their masses are expected to be more sensitive to the parameters affecting
rotational energy. Likewise, since the ratio $\lambda/\mu$ decreases for $X=$
$^{16}$O and $^{40}$Ca, the size of the nucleus also decreases with
$R_{0}=1.3943$ fm \ and $1.3470$ fm respectively.%

\[%
\begin{tabular}
[c]{|c|c|c|c|c|c|}\hline\hline
\multicolumn{6}{|c|}{Table I: Nuclear masses (MeV)}\\\hline\hline
$B$ & \ Nucleus\  &
\begin{tabular}
[c]{c}%
Set Ia\\
\multicolumn{1}{l}{$E_{B}($N+$^{4}$He$)$}%
\end{tabular}
&
\begin{tabular}
[c]{c}%
Set Ib\\
\multicolumn{1}{l}{$E_{B}($N+$^{16}$O$)$}%
\end{tabular}
&
\begin{tabular}
[c]{c}%
Set Ic\\
\multicolumn{1}{l}{$E_{B}($N+$^{40}$Ca$)$}%
\end{tabular}
& $\quad E_{exp}\quad$\\\hline\hline
$1$ & nucleon & $\rightarrow$ & $\rightarrow$ & $\rightarrow$ & $938.919$\\
$2$ & $^{2}$H & $1869.63$ & $1868.58$ & $1867.92$ & $1875.61$\\
$3$ & $^{3}$H & $2798.13$ & $2795.75$ & $2794.23$ & $2808.92$\\
$4$ & $^{4}$He & $\rightarrow$ & $3723.77$ & $3721.47$ & $3727.38$\\
$6$ & $^{6}$Li & $5592.02$ & $5586.73$ & $5583.36$ & $5601.52$\\
$7$ & $^{7}$Li & $6524.66$ & $6518.57$ & $6514.68$ & $6533.83$\\
$9$ & $^{9}$Li & $8387.75$ & $8379.78$ & $8374.70$ & $8392.75$\\
$10$ & $^{10}$B & $9320.89$ & $9312.17$ & $9306.62$ & $9324.44$\\
$16$ & $^{16}$O & $14909.5$ & $\rightarrow$ & $14885.9$ & $14895.1$\\
$20$ & $^{20}$Ne & $18636.9$ & $18618.8$ & $18607.3$ & $18617.7$\\
$40$ & $^{40}$Ca & $37273.8$ & $37237.7$ & $\rightarrow$ & $37214.7$\\
$56$ & $^{56}$Fe & $52183.4$ & $52132.9$ & $52100.7$ & $52089.8$\\
$238$ & $^{238}$U & $221780$ & $221565$ & $221429$ & $221696$\\\hline
\end{tabular}
\ \
\]
Given this unexpected success, one may wonder how switching on the nonlinear
$\sigma$ and Skyrme terms can improve or affect these results. Indeed the last
results suggest that these contributions need not be be very large. This
aspect is analysed in the next section.

\section{\label{sec:Approximation}Nonlinear $\sigma$ and Skyrme terms}

Let us now consider the full Lagrangian in (\ref{L_S}) assuming that the
contribution the nonlinear $\sigma$ and Skyrme terms can be set arbitrarily
small so that (\ref{Fsoln}) represents a good approximation to the exact
solution. Inserting the solution in (\ref{Estat}) and in the expression for
the various moments of inertia, one get additional contributions proportional
to $\alpha$ and $\beta$
\begin{align}
E_{stat}  &  =2n\pi\mu\lambda+16\pi\alpha\left(  \frac{n\lambda}{3\mu}\right)
^{\frac{1}{3}}\Gamma\left(  \frac{1}{3}\right)  \left(  \left(  2-2^{\frac
{2}{3}}\right)  \left(  n^{2}+1\right)  +2\zeta\left(  \frac{7}{3}\right)
\right) \nonumber\\
&  +\frac{128\pi\beta}{3}\left(  \frac{3\mu}{n\lambda}\right)  ^{\frac{1}{3}%
}\Gamma\left(  \frac{2}{3}\right)  \left(  \left(  8\left(  2\cdot3^{\frac
{1}{3}}-2^{\frac{2}{3}}\right)  -7\cdot2^{\frac{1}{3}}\right)  n^{2}%
+2^{\frac{1}{3}}\right)  \label{E4par}%
\end{align}
and
\begin{equation}
E_{rot}=\frac{1}{2}\left[  \frac{j(j+1)}{V_{11}}+\frac{i(i+1)}{U_{11}}+\left(
\frac{1}{U_{33}}-\frac{1}{U_{11}}-\frac{n^{2}}{V_{11}}\right)  \kappa
^{2}\right]  . \label{Ekappa2}%
\end{equation}
with $\kappa=0$ or $\frac{1}{2}$ for even and odd $B$ respectively and
\begin{align}
U_{11}  &  =64\pi\alpha\left(  \frac{n\lambda}{\mu}\right)  +\frac{512\pi
\beta}{9}\left(  \frac{3n\lambda}{\mu}\right)  ^{\frac{1}{3}}\Gamma\left(
\frac{1}{3}\right)  \left(  12^{\frac{1}{3}}+\left(  3n^{2}+1\right)  \left(
-4+6^{\frac{1}{3}}\left(  1+2^{\frac{1}{3}}\right)  \right)  \right)
\nonumber\\
&  +2\pi\left(  \frac{\lambda n}{3\mu}\right)  ^{\frac{5}{3}}\mu^{2}%
\Gamma\left(  \frac{2}{3}\right)  \left(  16\cdot3^{\frac{1}{3}}%
-9\cdot2^{\frac{2}{3}}\right)  \frac{\left(  3n^{2}+1\right)  }{4n^{2}}
\label{U4par}%
\end{align}%
\begin{align}
U_{33}  &  =64\pi\alpha\left(  \frac{n\lambda}{\mu}\right)  +\frac{512\pi
\beta}{9}\left(  \frac{3n\lambda}{\mu}\right)  ^{\frac{1}{3}}\Gamma\left(
\frac{1}{3}\right)  \left(  12^{\frac{1}{3}}+4\left(  -4+6^{\frac{1}{3}%
}\left(  1+2^{\frac{1}{3}}\right)  \right)  \right) \nonumber\\
&  +2\pi\left(  \frac{\lambda n}{3\mu}\right)  ^{\frac{5}{3}}\mu^{2}%
\Gamma\left(  \frac{2}{3}\right)  \left(  16\cdot3^{\frac{1}{3}}%
-9\cdot2^{\frac{2}{3}}\right)  \frac{1}{n^{2}} \label{U34par}%
\end{align}%
\begin{align}
V_{11}  &  =64\pi\alpha\left(  \frac{n\lambda}{\mu}\right)  \frac{\left(
n^{2}+3\right)  }{4}+\frac{128\pi\beta}{9}\left(  \frac{3n\lambda}{\mu
}\right)  ^{\frac{1}{3}}\Gamma\left(  \frac{1}{3}\right)  \left(  \left(
n^{2}+3\right)  12^{\frac{1}{3}}+16n^{2}\left(  -4+6^{\frac{1}{3}}\left(
1+2^{\frac{1}{3}}\right)  \right)  \right) \nonumber\\
&  +2\pi\left(  \frac{\lambda n}{3\mu}\right)  ^{\frac{5}{3}}\mu^{2}%
\Gamma\left(  \frac{2}{3}\right)  \left(  16\cdot3^{\frac{1}{3}}%
-9\cdot2^{\frac{2}{3}}\right)  \label{V4par}%
\end{align}
and as above $W_{11}=\delta_{n,1}U_{11}$ otherwise $W_{11}=W_{22}=0$ for
$\left\vert n\right\vert \geq2$ . Again due to the axial symmetry of the
ansatz, $U_{11}=U_{22}\neq U_{33}$ while non diagonal elements of $U_{ij}$ are
zero. Similar identities also holds for the $V_{ij}$ and $W_{ij}$ tensors.
Furthermore we have $n^{2}U_{33}=nW_{33}=V_{33}$. Relations (\ref{E4par}%
-\ref{V4par}) bring a clear understanding of the dependence of the masses of
the nuclei on the various parameters $B=n,$ $\mu$, $\alpha,$ $\beta$ and
$\lambda$ as long as $\alpha$ and $\beta$ remain relatively small.

In order to estimate the magnitude of the parameter $\alpha$ and $\beta$ in a
real physical case, we perform two more fits: Set II optimizes the four
parameters $\mu$, $\alpha,$ $\beta$ and $\lambda$ to reproduce the best fit
for the masses of the nuclei and Set III is done with respect to the ratio of
the binding energy over atomic number, $B.E./A$. More precisely, we use only a
subset of table of nuclei \cite{Nucltable} composed of the most abondant 144
isotopes (see Fig. 1). This is compared to Set I which was determined in the
previous section using the masses of the nucleon and $^{4}$He and assuming
$\alpha=\beta=0.$ The results are presented in Fig. 1 in the form of $B.E./A$
as a function of the baryon number for Sets Ia, II, III and experimental
values. The optimal values of the parameters are%

\[%
\begin{tabular}
[c]{|c|c|c|c|}\hline\hline
\multicolumn{4}{|c|}{Table II: Value of parameters for different
fits}\\\hline\hline
\ Nucleus\  &
\begin{tabular}
[c]{c}%
Set Ia\\
\multicolumn{1}{l}{(N+$^{4}$He$)$}%
\end{tabular}
&
\begin{tabular}
[c]{c}%
Set II\\
\multicolumn{1}{l}{(Masses)}%
\end{tabular}
&
\begin{tabular}
[c]{c}%
Set III\\
($B.E./A)$%
\end{tabular}
\\\hline\hline
$\mu$ $($MeV$^{2})$ & $30174.2$ & $29841.2$ & $29475.7$\\
$\alpha$ $($MeV$^{2})$ & $0$ & $0.00830341$ & $0.0316869$\\
$\beta$ $($dimensionless$)$ & $0$ & $-5.48285\times10^{-7}$ & $-4.01085\times
10^{-7}$\\
$\lambda$ $($MeV$^{-1})$ & $0.00491505$ & $0.00496265$ & $0.00503994$\\\hline
\end{tabular}
\ \ \ \ \
\]

As suspected the new sets of parameters are very close to Set Ia. . The
nonlinear $\sigma$ and Skyrme parameters $\alpha$ et $\beta$ are very small
but in order to compare, it is best to rescale the static energy with the
change of variable $u=\left(  4\mu/3\lambda\right)  ^{\frac{1}{3}}r$ such that
the relative weight of each term is more apparent$.$ Then the static energy
takes the form
\begin{align}
E_{stat}  &  =4\pi\left(  \frac{3\lambda\mu}{4}\right)  \int u^{2}du\left(
V+2\alpha\left(  \frac{4}{3\lambda\mu^{2}}\right)  ^{\frac{2}{3}}\left[
F^{\prime2}+\left(  n^{2}+1\right)  \frac{\sin^{2}F}{u^{2}}\right]  \right. \\
&  +\left.  16\beta\left(  \frac{16}{9\lambda^{2}\mu}\right)  ^{\frac{2}{3}%
}\frac{\sin^{2}F}{u^{2}}\left[  \left(  n^{2}+1\right)  F^{\prime2}+n^{2}%
\frac{\sin^{2}F}{r^{2}}\right]  +n^{2}F^{\prime}{}^{2}\frac{\sin^{4}F}{u^{4}%
}\right)
\end{align}
\begin{figure}[ptbh]
\centering\includegraphics[width=0.7\textwidth]{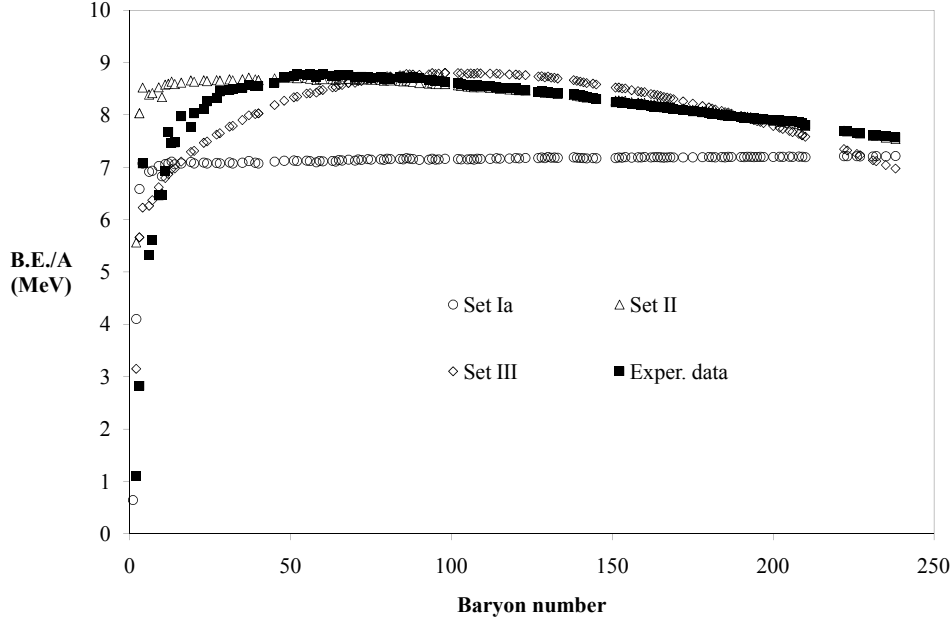} \caption{Ratio of
the binding energy ($B.E.$) over the atomic number $A$ (or baryon number) as a
function of $A$. The experimental data (black squares) are shown along with
predicted value for parametization of Set Ia (empty circles), II (empty
triangles) and III (empty diamonds) respectively.}%
\label{fig1}%
\end{figure}where $F^{\prime}=\partial F/\partial u$ and the energy can be
expressed in units of $\frac{3\lambda\mu}{4}$. For example for Set II (Set
III), the nonlinear $\sigma$ term is proportional to $\alpha\left(  \frac
{4}{3\lambda\mu^{2}}\right)  ^{\frac{2}{3}}=3.73524\times10^{-7}$
$(1.43418\times10^{-6})$ and the Skyrme term to $\beta\left(  \frac
{16}{9\lambda^{2}\mu}\right)  ^{\frac{2}{3}}=-1.3263\times10^{-6}$
$(-9.2355\times10^{-6})$ while the remaining terms are of order one.
Furthermore the overall factor $\left(  \frac{3\lambda\mu}{4}\right)  $
remains approximately the same for all the sets. Looking at the numerical
results, we observe nonetheless that these two terms are responsible for
corrections of the order of $0.01\%.$ Clearly, the small magnitude of these
contributions provides support to the assumption that (\ref{Fsoln}) is a good
approximation to the exact solution.

Comparing Set II to the original Skyrme Model with a pion mass term, we may
identify
\begin{align*}
F_{\pi}  &  =4\sqrt{\alpha}=0.364474\text{ MeV\qquad(Experiment: }F_{\pi
}=\text{186 MeV)}\\
e^{2}  &  =\frac{1}{32\beta}=-57019\qquad\text{(}e=4.84\text{ for massive pion
Skyrme Model)}\\
m_{\pi}  &  =\frac{2\sqrt{\alpha}}{\mu}=231591\text{ MeV\qquad(Experiment:
}m_{\pi}=\text{138 MeV)}%
\end{align*}
Set III leads to similar values for $F_{\pi},e^{2}$ and $m_{\pi}$ which are
orders of magnitude away for the usual values obtained for the Skyrme model.
Of course here the nonlinear $\sigma$ and Skyrme terms do not play a
significant role in the stabilization of the soliton. Indeed the Skyrme term
even have the wrong sign so it would destabilize the soliton against shrinking
if it was not for the contribution of order six in derivatives. The size of
the soliton is instead determined by the relative magnitude of $\mu$ and
$\lambda$ so there is no need for $F_{\pi}$ and $m_{\pi}$ to be close to the
nucleon mass scale as for the original Skyrme Model. Perhaps the explanation
for such a departure is that the parameters of the model are merely bare
parameters and they could differ significantly from their renormalized
physical values. We note also that the $B>1$ solutions of the Skyrme Model
display a totally different structure compared to the BPS-type solution
analyzed here. It is well known that the lowest-energy $B=2,3,4...$ solutions
of the Skyrme Model exhibit respectively toroidal, tetrahedral, cubic,...
baryon density configurations. Such solution are conveniently represented by
an ansatz based on rational maps \cite{Houghton}. The model at hand here leads
to spherically symmetric baryon density at least in the regime of small
$\alpha$ and $\beta$ where solution (\ref{Fsoln}) can apply. So it seems that
the regime dominated by the $\mu$ and $\lambda$ terms leads to spherical
configurations whereas the regime dominated by the nonlinear $\sigma$ and
Skyrme terms shows totally different baryon density distributions. In the
absence of a complete analysis, we can only conjecture that the change in
configuration is related to which of the four terms are responsible for the
stabilization of the soliton and at some critical values of the parameters
there is a transition between configurations.

Let us now look more closely at the numerical results presented in Fig. 1.
These are in the form of the ratio of the binding energy ($B.E.$) over the
atomic number $A$ as a function of $A$ which corresponds to the baryon number.
The experimental data (black squares) are shown along with predicted value for
parametization of Set Ia (empty circles), II (empty triangles) and III (empty
diamonds) respectively. Clearly Set Ia is less accurate when it comes to
reproduce the full set of experimental data but is somewhat successful for the
lightest nuclei. This to be expected since the fit relies on the masses of the
nucleon and $^{4}$He. Yet all predicted nuclear masses are found to be within
a 0.3\% precision. In fact the ratio $B.E./A$ is rather sensitive to small
variation of the nuclear masses so the results in general are surprisingly
accurate. On the other hand Set II, based on the nuclear masses, overestimates
the binding energies of the lightest nuclei while it reproduces almost exactly
the remaining experimental values. The least square fit based on $B.E./A$, Set
III, is the best fit overall but in order to better represent the features of
lightest nuclei, it abdicates some of the accuracy found in Set II for $B>30$.

This apparent dichotomy between the description of the two regions $B\leq30$
and $B>30$ may find an explanation in the (iso)rotational contribution to the
mass. Indeed light nuclei have smaller sizes and moments of inertia so that
their rotational energy contributes to a larger fraction of the total mass
since the spins and isospins remain relatively small. On the other hand the
size of heavy nuclei grows as $B^{\frac{1}{3}}$ and their moments of inertia
increase accordingly. The spin of the most abundant isotopes are relatively
small while isospin can have relatively large values due to the growing
disequilibrium between the number of proton and the number of neutron in heavy
nuclei (see eq. \ref{Isospin}). Despite these behaviors, our numerical results
show that the rotational energy is less that 1 MeV for $B>10$ for any of the
Sets considered and its contribution to $B.E./A$ decreases rapidly as $B$
increases. On the contrary for $B<10$ the rotational energy is responsible for
large part of the binding energy which means that $B.E./A$ \ should be very
sensitive to the way the rotational energy is computed. In our case we
approximated the nucleus as a rigid rotator. One may argue that if rotational
deformations due to centrifugal effects were to be considered, it would lead
to larger moments of inertia and lower rotational energies. This would
predominantly affect the binding energy of the lightest nuclei since this is
where rotational energy is most significant. Allowing for such deformation
would in general require the full numerical computation of the solution. An
easier way to check for deformation is by allowing the ratio of the parameters
$\sigma=\mu/\lambda$ in the solution (\ref{Fsoln}) to vary independently from
the $\mu$ and $\lambda$ in the model (\ref{L_S}) and by repeating the fit with
respect to five parameters instead of the four previous ones. This procedure
allows for a further adjustment of the size of the soliton in terms of
$\sigma$ with respect to a given choice of model parameters $\mu,\alpha,\beta$
and $\lambda$ and would lead to partial deformation of the solution. Such a
parametrization is expected to increase both the size and the moments of
inertia of the soliton and decrease the total mass of the lightest nuclei
which would be an improvement over the four parameters fit. We evaluated such
correction for the nucleon whose relative contribution to mass from rotation
is the largest using the parameters of Set II and we obtained a modest
decrease of the mass of the order of 0.16\%. Since the rotational energy
accounts for much less than 1\% of the total energy in most of the nuclei,
deformations are not generally expected to be very significant.

\section{Conclusion}

We have proposed a 4-terms model as a generalization of the Skyrme Model. In
the regime where two of the terms are negligible, i.e. $\alpha=\beta=0$, we
find well-behaved analytical solutions for the static solitons. These saturate
the Bogomolny's bound with consequence that the static energy is directly
proportional to the baryon number $B$. They differ from those obtained by ASW
in an important way: their model lead to compactons at the boundary of which
the gradient of the solution is infinite and so the solution could not be used
to approximate the energies in the regime where $\alpha,\beta\neq0$.
Furthermore, one of the major feature of our model is that the form of the
solutions allows to compute analytically the static and rotational energy and
express them as a function of the model parameters and $B$. Fixing the
remaining parameters of the model, $\mu$ and $\lambda,$ leads to rather
accurate predictions for the mass of the nuclei.

We then used these BPS-type solutions to compute the mass of the nuclei in the
regime where $\alpha$ and $\beta$ are small but not zero. Indeed fitting the
model parameters to provide the best description of the nuclear mass data
leads to that particular regime where the values of $\alpha$ and $\beta$ turn
out to be very small. Yet we find a noticeable improvement in the size and
$B.E./A$ predictions with respect to those for the $\alpha=\beta=0$ regime.
Even though our 4-term model can be considered a simple extension of the
massive pion Skyrme Model (different mass term and an additional term with six
derivatives in pion fields) the solution leads to spherically symmetric baryon
densities as opposed to more complex configurations for $B>1$ standard
Skyrmions (e.g. toroidal, tetrahedral, cubic...). These results suggest that
nuclei could be considered as near BPS-Skyrmions.

This work was supported by the National Science and Engineering Research
Council of Canada.

\end{document}